\documentclass[12 pt]{article}
\usepackage{eurosym}
\usepackage{amssymb}
\usepackage{amsmath}
\usepackage{amsfonts}
\usepackage{graphicx}
\usepackage{geometry}
\usepackage{array}
\usepackage{xcolor}
\usepackage{tikz}
\usepackage{colortbl}
\usepackage{graphicx}

\usetikzlibrary{arrows.meta,automata,positioning}
\newtheorem{theorem}{Theorem}

\newtheorem{corollary}[theorem]{Corollary}

\newtheorem{example}{Example}

\newtheorem{proposition}[theorem]{Proposition}

\newenvironment{proof}[1][Proof]{\noindent\textbf{#1.} }{\ \rule{0.5em}{0.5em}}

\begin{document}

\title{Applying an axiomatic approach to revenue allocation in airlines
problems\thanks{%
Financial support from grant PID2020-113440GBI00 funded by MCIN/AEI/
10.13039/501100011033 is gratefully acknowledged. The authors also
acknowledge the financial support of the Universidade de Vigo through its
Contract-Program funding scheme.}}
\author{Gustavo Berganti\~nos \\
%EndAName
ECOBAS, Universidade de Vigo, ECOSOT, 36310 Vigo, Espa\~{n}a;
gbergant@uvigo.es\\
Leticia Lorenzo\\
Economics, Society and Territory.\\
CITMAga.\\
Universidade de Vigo.}
\date{}
\maketitle

\begin{abstract}
The International Air Transport Association (IATA) states that the revenue
from interline tickets must be shared among the different airlines according
to a weighted system. We analyze this problem following an axiomatic
approach, and our theoretical results support IATA's procedure. Our first
result justifies the use of a weighted system, but it does not specify which
weights should be applied. Assuming that the weights are fixed, we provide
several results that further support the use of IATA's mechanism. Finally,
we provide results for the case in which all flights can be considered
equivalent and no weighting is required.

\end{abstract}

\section{Introduction}

In the aviation industry, airlines frequently form alliances to improve
their connectivity, optimize operational costs, and provide better services
to passengers. This collaboration through resource sharing improves
efficiency and leads to an increase in their overall revenues. However, the
way in which the revenues are distributed among the different members of the
alliance is fundamental to ensuring the alliance remains stable and lasts
over time.

In this paper, we address a problem of revenue allocation among airlines
inspired by the International Air Transport Association (IATA), which
employs a proration system to fairly distribute revenues among airlines when
a single itinerary involves multiple flights. The Multilateral Proration
Agreement - Passenger (MPA-P)\footnote{%
See https://www.iata.org/en/services/finance/prorate-manuals/} is a
contractual framework that defines how revenue from interline tickets must
be shared among the different airlines. IATA provides the formulas, data,
and procedures necessary for prorating passenger revenues between airlines
according to the MPA-P.

This proration system is essential to guarantee equitable compensation based
on each airline's contribution to the overall journey. The proration
mechanism is fundamentally based on the distances flown by each airline.
However, to make this distribution fairer, correction factors are applied to
those distances. These factors are determined by considering various
variables such as operating costs, regional differences, and so on. These
corrections help ensure that the proportional distribution does not solely
favor long-haul routes but also includes adjustments for the specific
characteristics of each route.

We first consider a theoretical model that is built on a graph
representation where nodes correspond to airports, and edges represent
flights or legs. Each flight is defined by its origin, destination, and
operating airline. We also consider the possibility that the flight has an
associated weight that may represent certain characteristics of the flight,
for instance, the proration factor used by IATA. Besides, we incorporate
data on passenger itineraries, detailing the routes taken, the airlines
involved, and the fares paid. The objective is to allocate the total revenue
generated from these passenger itineraries among the airlines in a fair way.
In our model, IATA's revenue allocation is as follows: the
amount paid by each passenger is divided among the airlines operating the
passenger's itinerary in proportion to the total weight of
the flights operated by each airline. The amount received by each airline is
computed by adding the amounts received from all passengers.

We study this problem following an axiomatic approach. Thus, we consider
several axioms that formalize normatively appealing principles. By applying
some of those axioms, we obtain axiomatic characterizations of families of
rules or single rules.

IATA's revenue allocation depends on the weights computed through MPA-P. One
may consider allocations using IATA's procedure but based on other weights
(for instance, distances between airports). Thus, we consider the family of
all weighted flights rules, where for each possible weight system we compute
the associated revenue allocation. Our first result is an axiomatic
characterization of the family of weighted flights rules based on four
axioms. The first axiom is the classical additivity axiom (the rule is
additive on the set of passengers). The second axiom is also classical: null
airline (if an airline has no passengers, it receives nothing). The third
axiom, independence of other airlines, is new and says that the amount
received by each airline depends on the flights it operates, but not on
which airlines operate the remaining flights. The fourth axiom, ratio
preservation, is also new and says that, if in the itinerary of two
passengers, airlines $i$ and $i^\prime$ operate the same flights, and
airline $i$ receives $\lambda $ times more than airline $i^{\prime }$ from
the first passenger, then airline $i$ should also receive $\lambda $ times
more than airline $i^{\prime }$ from the second passenger. This result
partially supports the revenue allocation used by IATA. If we want to apply
a rule satisfying the previous four axioms, then we should use a weighted
flights rule, as IATA does. Nevertheless, the axioms do not determine which
specific weights should be used.

It is clear that some flights are very different (for instance, domestic and
international flights) and so should be treated differently. Besides, we
assume that it is possible to compare the importance of different flights
through a specific weight system, as IATA does. Under such circumstances, it
seems natural to incorporate these weights as part of the theoretical model.
We do so and, besides, we consider an additional axiom that relates
explicitly to the weights. Pairwise homogeneity says that if, for each
passenger, the weights of the flights operated by two airlines are in a
given proportion, the total benefit received by these two airlines should
respect the same proportion. We provide three characterizations of the
weighted flights rule associated with the given weights. Two axioms
(additivity and pairwise homogeneity) appear in the three characterizations.
The other three axioms: null airline, independence of empty flights (if we
cancel a flight with no passengers, no airline is affected), and
independence of other airlines, each appear in only one. These axiomatic
characterizations also support the revenue allocation used by IATA, assuming
that the weights adequately capture the asymmetries among flights.

The weights assigned by IATA to each flight are rather arbitrary (for
instance, they are updated annually). In some cases, flights do not differ
much (for instance, in Spain, most domestic flights have a duration of
around one hour). Thus, in such situations, it makes sense to assume that
all flights are equal, or equivalently, that they all have the same weight.
We admit that this is not a good general philosophy, but it could be
reasonable in certain specific contexts. In these cases, we could divide the
revenue by following the so-called equal flights rule, under which the
amount paid by each passenger is divided among the airlines proportionally
to the number of flights in the passenger's itinerary
operated by each airline. We provide an axiomatic characterization of this
rule based on additivity, independence of other airlines, and flights
equivalence (if all passengers have booked the same number of flights from
two airlines, then those two airlines must receive the same).

\bigskip

\subsection{Related literature}

In the aviation industry, many issues have been analyzed using game theory.
Recently, Dixit $et$ $al$ (2024) considered a model with one airport and two
airlines, in which they studied how reciprocity and fairness impact on
environmental sustainability, and Xu $et$ $al$ (2024) proposed a model to
jointly compute the flight schedule, fleet assignment, and airfares.
Littlechild and Owen (1973) studied how to set landing fees at an airport.
Houghtalen $et$ $al$ (2011) designed mechanisms for sharing profits among
the members of an alliance in such a way that the final result is optimal
for the alliance as a whole. Hu $et$ $al$ (2013) proposed a two-stage game
for sharing the revenues in an airline alliance: in the first stage,
airlines negotiate proration rates for the revenues generated by shared
itineraries; in the second stage, airlines operate independently trying to
maximize their own revenues. They consider a revenue-sharing rule for the
first stage that induces an optimal result for the alliance as a whole.

Kimms and Cetiner (2012) and Wang (2020) studied the same problem as we do,
but with some significant differences. Kimms and Cetiner (2012) defined a
cooperative game with transferable utility, and they proposed to share the
revenues using the nucleolus of this game. Wang (2020) also defined a
cooperative game, different from that of Kimms and Cetiner (2012), and
proposed to share the revenues using the Shapley value of this game.
Besides, he gave an axiomatic characterization of the Shapley value and thus
also adopted an axiomatic approach, as we do in this paper.

We now briefly compare our paper with Wang (2020). The graph in our model is
arbitrary, whereas the graph in Wang (2020) is a tree. This implies, for
instance, that if there is a flight from Madrid to Paris and another from
Paris to London, then there cannot be a direct flight from Madrid to London,
which is an assumption that is not realistic. In Wang (2020), only one
airline can operate the flights between any two airports, whereas we allow
multiple airlines to do so. In many real cases, several airlines operate
flights between the same pair of airports. Thus, our model can accommodate
all real situations, whereas Wang's (2020) model is rather
limited. The Shapley rule considered in Wang (2020) is different from the
rules considered in our paper. The additivity and null airline axioms used
here also appear in Wang (2020), under the names of separability and null
airline property respectively. Wang (2020) also uses the axiom of equal
treatment of equals, which we do not use in this paper.

Our paper is directly connected to the axiomatic literature on resource
allocation (e.g., Thomson, 2023). Some recent examples of this literature
are the following: Martinez and Moreno-Ternero (2022) characterized a family
of pandemic indicators computed as a weighted average of the incidence,
morbidity, and mortality rates. Gudmundsson $et$ $al$ (2024) characterized a
class of fixed-fraction rules, which balance between incentives for accident
prevention and fairness, to assign liabilities in the problem of sharing
sequentially triggered losses. Finally, Berganti\~{n}os and Moreno-Ternero
(2025a,b) and Gon\c{c}alves-Dosantos $et$ $al$ (2025a,b) studied how to pay
content creators on streaming platforms.

\bigskip

The paper is structured as follows. Section 2 summarizes the proration
system applied by IATA; Section 3 introduces the model and the proposed
rules; Section 4 introduces the different axioms; Section 5 provides
axiomatic characterizations of the family of all weighted flights rules;
Section 6 provides axiomatic characterizations of the weighted flights rule;
Section 7 provides an axiomatic characterization of the equal flights rule;
and finally, Section 8 sets out our conclusions.

\bigskip

\section{IATA's Prorate system}

IATA has a proration mechanism to divide the revenue associated with a
passenger's journey where multiple airlines, within the same alliance, are
included. This procedure specifies how the revenue is distributed according
to certain established guidelines. What is described in this section is a
summary intended to briefly outline how the prorate system works without
delving into the many details and exceptions. For more detailed information,
we refer to the MPA-P.

The following steps are taken into account in order to compute the
allocation of benefits:

\begin{enumerate}
\item Compute the amount to be prorated (ATBP). This amount is not the one
specified on the passenger's ticket, as certain components must be excluded.
In particular, taxes, fees, and other charges that are not part of the base
fare must be removed. For example, if a passenger pays an extra fee to
travel in first class, that surcharge should be allocated to the airline
operating that specific leg, while the base fare would be included in the
ATBP.

\item Establish the Prorate Factors in Terms of Mileage. These factors are
based on Ticketed Point Mileages (TPM). Although TPMs, published annually by
IATA, are very close to the geodesic distance, they do not match exactly, as
TPMs are calculated for fare-related purposes and may incorporate other
commercial or political considerations.\footnote{%
For the sake of readability, we will refer to TPM and distance
interchangeably throughout the paper.}

These TPMs are multiplied by what is known as the Worldwide Weight,
resulting in the Adjusted Proration Factors. IATA uses the following
formula, applicable from March 2025\footnote{%
The formula is updated annually, taking into account how operating costs per
seat and per mile vary with the distance traveled. The computation uses the
data provided by the International Civil Aviation Organization (ICAO).}, in
order to compute the Worldwide Weight: 
\begin{equation*}
\text{ Worldwide Weight }=6.338826\cdot \text{TPM}^{-0.209892}
\end{equation*}
It is interesting to note that as the distance increases, the weight
decreases. What we observe is that longer segments are somehow penalized
since shorter segments tend to be more expensive per mile due to takeoffs,
landings, and associated fees.

Finally, the adjusted proration factors are further modified by applying
additional weights. Since operational costs, market conditions, and regional
factors vary, IATA applies various correction coefficients to adjusted
prorate factors depending on the regions involved in the itinerary. IATA
specifies several weights according to the origin and destination of the
flight, which are also updated annually. These adjustments ensure fair
compensation for airlines operating in high-cost or less competitive regions.

After applying these weightings, each factor is rounded to the nearest whole
number, producing the Standard Proration Factor (SPF).

\item Compute the allocation of the revenue among the airlines. Once the
SPFs are computed, the ATBP is allocated proportionally to those factors.
\end{enumerate}

\begin{example}
Let us consider a passenger who travels from Madrid (MAD) to Nairobi (NBO)
through Frankfurt (FRA). The first flight is operated by Airline 1 and the
second one by Airline 2. The AMTP is 900 USD. First we compute the adjusted
prorate factors taking into account the distance between each place (we
consider geodesic distance in the example).\medskip

Airline-adjusted prorate factors for the itinerary:

\begin{tabular}{lccc}
Segment & TPM & Worldwide weight & Adjusted TPM \\ \hline
MAD - FRA & $893$ & $1.52$ & $893\times 1.52=1339.5$ \\ 
FRA - NBO & $3690$ & $1.13$ & $3690\times 1.13=4169.7$%
\end{tabular}

Regional adjustments are applied based on the region of origin and
destination for each segment. In the case of the first flight within Europe,
under the latest proration agreement, that factor would be 0.97. The weight
for flights between Europe and Africa is 1.142. Thus, the final factors are:

\begin{tabular}{lcc}
Segment & Adjusted TPM & Standard Prorate Factor (SPF) \\ \hline
MAD - FRA & $1339.5$ & $round(1339.5\times 0.97)=1299$ \\ 
FRA - NBO & $4169.7$ & $round(4169.7\times 1.142)=4760$%
\end{tabular}

The final allocation is computed proportionally to the SPF as follows: 
\begin{align*}
\text{Airline 1: }& \frac{1299}{1299+4760}\cdot 900=192.95\text{ }USD \\
\text{Airline 2: }& \frac{4760}{1299+4760}\cdot 900=707.05\text{ }USD
\end{align*}

Note that applying the TPMs directly, without incorporating the different
weights, will lead to the allocation $(175.36, 724.63)$, which clearly
benefits the airline operating the longer segment.
\end{example}

\bigskip

\section{The model}

In this section, we introduce the mathematical model intended to represent
the situation under study. This setting involves several aspects that must
be considered: the flight offerings of the various airlines, the itineraries
completed by a group of passengers, and the benefits associated with those
itineraries, which are to be distributed among the airlines. \medskip

The flight offering is modeled as a directed graph, $g=\left( V,E\right) $,
where $V$ is the set of vertices and $E\subset V\times V$ is the set of
edges. Each vertex represents an airport and each directed edge $%
\left(i,j\right) \in E$ with $i\neq j$ represents a direct flight from $i$
to $j$.

Let $\mathcal{N}\subset \{1,2,\ldots\}$ be the set of all possible airlines.
We consider a finite subset $N\subset \mathcal{N}$ of airlines. Each airline
is characterized by the flights it operates.

We denote the available flights as $F\subset E\times N$. So, a flight $%
(e,i)\in F$ is characterized by its origin and destination $e\in E$, and the
airline operating it $i\in N$.

It may occur that several airlines offer different flights connecting the
same pair of airports. That is, given an arc $e$ and two different airlines $%
i,j\in N$, it is possible that $(e,i),(e,j)\in F$. For simplicity, we do not
allow multiple arcs between the same pair of airports operated by the same
airline. We know that it is quite common for a given airline to offer
several flights between two airports. Our model also covers this situation.
In our model, an arc represents the possibility of traveling between two
airports, but there may be multiple schedule options for the same flight,
each with different prices. That is, two different passengers could fly
directly from the same origin to the same destination with the same airline,
but one in the morning and the other in the afternoon, and, as occurs in
reality, pay different prices for their tickets. If we want to consider
several distinct flights (arcs), we would need to introduce additional
notation. This would make the model unnecessarily complex, because the
results would remain essentially the same.

We also assume that each pair of airports is connected by flights. As far as
we know this assumption is realistic. We say that $a$, $b\in V$ are directly
connected in $F$ when there is a direct flight from $a$ to $b.$ Namely,
there exists $i\in N$ such that $\left( \left( a,b\right) ,i\right) \in F.$
We say that $a$, $b\in V$ are connected in $F$ when there is a path in $F$
from $a$ to $b$. Namely, a sequence $\left\{ \left(
\left(a_{k},a_{k+1}\right) ,i^{k}\right) \right\} _{k=1}^{l}$ such that $%
a_{1}=a$, $a_{l+1}=b$ and for each $k=1,\ldots ,l$, $\left(
\left(a_{k},a_{k+1}\right) ,i^{k}\right) \in F$. We assume that for each $a$%
, $b\in V$, $a$ and $b$ are connected in $F.$

Let $E_{i}$ denote the set of flights operated by airline $i$. For each $%
i\in N$, 
\begin{equation*}
E_{i}=\{e\in E,(e,i)\in F\}.
\end{equation*}%
We assume that $E_{i}\neq \emptyset $; that is, we only consider airlines
operating at least one flight. \medskip

Additionally, we include passenger information in the following way. Let $M$
be the set of passengers. Each passenger $j\in M$ travels from an origin $%
o_{j}$ to a destination $d_{j}$, either directly or indirectly. Thus, each
passenger $j$ is characterized by a tuple $(f^{j},p^{j})$ where $%
f^{j}=\left\{ \left( e_{k}^{j},i_{k}^{j}\right)
\right\}_{k=1}^{n^{j}}\subseteq F$ is the set of flights taken by passenger $%
j$, $E^{j}=\left\{ e_{k}^{j}\right\} _{k=1}^{n^{j}}$ is a path in $E$ from $%
o_{j}$ to $d_{j}$, and $p^{j}>0$ is the price paid by passenger $j$\footnote{%
Throughout this paper, we will refer to the price paid by the passenger. In
reality, this price corresponds to the ATBP (Amount To Be Prorated), once
fees, taxes, and other charges have been excluded.}. We denote by $N^{j}$
the set of airlines operating the flights of passenger $j$. That is, 
\begin{equation*}
N^{j}=\left\{ i\in N:i=i_{k}\text{ for some }k=1,...,n^{j}\right\}
\end{equation*}

Besides, we can define the set of passengers that take a flight operated by
airline $i$ as 
\begin{equation*}
M_i=\{j\in M:(e,i)\in f^j \text{ for some }e\in E_i\}.
\end{equation*}

The set of flights offered by airline $i$ is 
\begin{equation*}
f_i=\{(e,i)\in F\}.
\end{equation*}

Finally, the set of flights in the journey of passenger $j$ operated by
airline $i$ is given by 
\begin{equation*}
f_i^j=\{(e,k)\in f^j:k=i\} .
\end{equation*}

\bigskip

An \textbf{airlines problem} is a tuple 
\begin{equation*}
A=\left( (V,E),N,F,M,(f^{j},p^{j})_{j\in M}\right) .
\end{equation*}

Since $(V,E)$ is fixed, we usually write 
\begin{equation*}
A=(N,F,M,(f^{j},p^{j})_{j\in M})
\end{equation*}

We denote by $\mathcal{P}$ the set of airlines problems.

\bigskip

A \textbf{weight system} is a vector $w=\left( w_{e}\right) _{e\in E}$ where
for each $e\in E,$ $w_{e}>0$ is the weight associated with flight $e$. Note
that the weight depends only on the origin and destination of the flight,
and not on the airline operating it. The weight $w_{e}$ may represent, for
example, the distance (miles), flight time, the IATA prorate factor, etc.

Sometimes the problem $A$ has an associated weight system $w$, and sometimes
it does not. In order to simplify the notation, we never include the weight
system explicitly in the definition of the problem. Nevertheless, the axiom
of pairwise homogeneity depends on this weight system. Thus, in our results
involving pairwise homogeneity, we assume that a weight system exists in $A$.

\bigskip

We now present an example to clarify the model introduced above.

\begin{example}
Consider an airlines problem with the flights represented in the graph. The
number associated with each arch indicates the airline operating that
flight. The value in parentheses indicates the weight $w$.

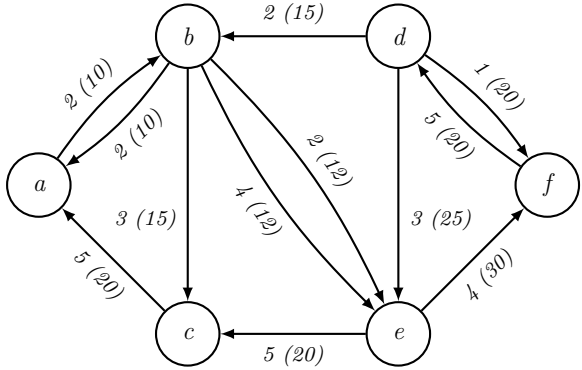
\begin{tikzpicture}[->,>=latex,auto,node distance=3.5cm,thick]
				\tikzstyle{every state}=[draw=black,thick,scale=0.8]
				\node[state](1){a};
				\node[state](2)[above right of=1]{b};
				\node[state](3)[below right of=1]{c};
				\node[state](4)[right of=2]{d};
				\node[state](5)[right of=3]{e};
				\node[state](6)[below right of=4]{f};
				
				\path[bend left=10] (1) edge  [sloped] node [above] {\scriptsize 2 (10)}(2);
				\path[bend left=10] (2) edge  [sloped] node [below] {\scriptsize 2 (10)}(1);
				\path (3) edge  [sloped] node [below] {\scriptsize 5 (20)}(1);
				\path (4) edge  [left] node [above] {\scriptsize 2 (15)}(2);
				\path[bend left=10] (2) edge  [sloped] node [above] {\scriptsize 2 (12)}(5);
				\path[bend right=10] (2) edge  [sloped] node [below] {\scriptsize 4 (12)}(5);
				\path (2) edge  [left,pos=0.65] node [left] {\scriptsize 3 (15)}(3);
				\path (5) edge  [right] node [below] {\scriptsize 5 (20)}(3);
				\path[bend left=10] (4) edge  [sloped] node [above] {\scriptsize 1 (20)}(6);
				\path[bend left=10] (6) edge  [sloped] node [below] {\scriptsize 5 (20)}(4);
				\path (4) edge  [right,pos=0.65] node [right] {\scriptsize 3 (25)}(5);
				\path (5) edge  [sloped] node [below] {\scriptsize 4 (30)}(6);
			\end{tikzpicture}

The airlines problem $A=(N,F,M,(f^{j},p^{j})_{j\in M})$ is given by:

\begin{itemize}
\item $N=\left\{ 1,2,3,4,5\right\} $. 

\item $F=%
\{((a,b),2),((b,a),2),((b,c),3),((b,e),2),((b,e),4),((c,a),5),((d,b),2),((d,e),3),((d,f),1), \linebreak((e,c),5),((e,f),4),((f,d),5)\} 
$. 

\item $M=\{1,2,3,4\}$. 

\item $(f^{j},p^{j})_{j\in M}$ is given by the following table:

\begin{tabular}{c|c|c}
Passenger $\left( j\right) $ & Journey $\left( f^{j}\right) $ & Price $%
\left( p^{j}\right) $ \\ \hline
1 & $f^{1}=\{((a,b),2)((b,e),4),((e,c),5)\}$ & $p^{1}=12$ \\ 
2 & $f^{2}=\{((b,e),2),((e,f),4),((f,d),5)\}$ & $p^{2}=30$ \\ 
3 & $f^{3}=\{((d,e),3),((e,c),5),((c,a),5)\}$ & $p^{3}=24$ \\ 
4 & $f^{4}=\{((a,b),2),((b,c),3)\}$ & $p^{4}=15$%
\end{tabular}
\end{itemize}

This example helps us to note that:

\begin{itemize}
\item Several airlines can operate the same flight. That is the case of the
flight from b to e, which is offered by airlines 2 and 4.

\item Passengers can have flights in common in their journeys. That is the
case of passengers 1 and 3, who booked the flight $((e,c),5)$. As explained
earlier, these passengers may have flown at different times

\item There may be flights offered by airlines that are not used by any
passenger. We can interpret these as empty flights. That is the case of
flights $((b,a),2)$, $((d,b),2)$ and $((d,f),1)$.
\end{itemize}
\end{example}

\bigskip

A \textbf{rule} $R$ is a function that divides the total amount paid by the
passengers among the airlines. Namely, given a problem $A$, $R(A)\in \mathbb{%
R}^{N}$ satisfies that $\sum_{i\in N}R_{i}(A)=\sum_{j\in M}p^{j}$ and $%
R_{i}(A)\geq 0$ for all $i\in N$.

\bigskip

We first present two possible rules. In both cases, profits are distributed,
on a passenger-by-passenger basis, among the airlines involved in each
individual journey. The final allocation is obtained by aggregating the
distributions across all passengers.\medskip

\bigskip

The first rule, called the \textbf{weighted flights rule}, follows the same
logic as the IATA's procedure. Given a weight system $w=\left(
w_{e}\right)_{e\in E},$ the amount paid by each passenger $j$ is divided
among the airlines proportionally to the weights of the flights of passenger 
$j$ operated by each airline. Formally, for each problem $A$ and each
airline $i\in N.$ 
\begin{equation*}
W_{i}^{w}(A)=\sum_{j\in M}\frac{\sum\limits_{(e,i)\in f_{i}^{j}}w_{e}}{%
\sum\limits_{(e,k)\in f^{j}}w_{e}}p^{j}.
\end{equation*}

When no confusion arises we write $W$ instead of $W^w$.

\bigskip

IATA uses this rule when the weights are computed according to the
Multilateral Proration Agreement - Passenger; that is, when the weights
correspond to the Standard Proration Factors.

The weights associated with each flight are rather arbitrary. For instance,
IATA updates them each year. Thus, if the flights are not very different, it
makes sense to consider that all of them are equal, or equivalently, they
have the same weight. We admit that this is not a good general philosophy
(for instance in the case of the flights MAD-FRA-NAI in Example 2).
Nevertheless, it could be reasonable in certain specific cases. For
instance, in Spain all domestic flights have a duration of around 1 hour. In
the \textbf{equal flights rule}, the amount paid by each passenger $j$ is
divided among the airlines proportionally to the number of flights of
passenger $j$ operated by each airline. Formally, for each problem $A$ and
each airline $i\in N$. 
\begin{equation*}
E_{i}(A)=\sum_{j\in M}\frac{|f_{i}^{j}|}{|f^{j}|}p^{j}.
\end{equation*}

Note that the equal flights rule coincides with any weighted flights rule
where all weights are equal (namely $w_{e}=w$ for all $e\in E)$.

\bigskip

We compute both allocation rules in Example 2.

\begin{example}
First we compute the weighted flights rule.

\begin{tabular}{c|c|c|c|c|c}
Airline & Passenger 1 & Passenger 2 & Passenger 3 & Passenger 4 & $%
W_{i}\left( A\right) $ \\ \hline
1 & $0$ & $0$ & $0$ & $0$ & $0$ \\ 
2 & $\frac{10}{42}12$ & $\frac{12}{62}30$ & $0$ & $\frac{10}{25}15$ & $%
14.6636$ \\ 
3 & $0$ & $0$ & $\frac{25}{65}24$ & $\frac{15}{25}15$ & $18.2308$ \\ 
4 & $\frac{12}{42}12$ & $\frac{30}{62}30$ & $0$ & $0$ & $17.9447$ \\ 
5 & $\frac{20}{42}12$ & $\frac{20}{62}30$ & $\frac{40}{65}24$ & $0$ & $%
30.1609$%
\end{tabular}

Next we compute the equal flights rule.

\begin{tabular}{c|c|c|c|c|c}
Airline & Passenger 1 & Passenger 2 & Passenger 3 & Passenger 4 & $%
E_{i}\left( A\right) $ \\ \hline
1 & $0$ & $0$ & $0$ & $0$ & $0$ \\ 
2 & $\frac{1}{3}12$ & $\frac{1}{3}30$ & $0$ & $\frac{1}{2}15$ & $21.5$ \\ 
3 & $0$ & $0$ & $\frac{1}{3}24$ & $\frac{1}{2}15$ & $15.5$ \\ 
4 & $\frac{1}{3}12$ & $\frac{1}{3}30$ & $0$ & $0$ & $14$ \\ 
5 & $\frac{1}{3}12$ & $\frac{1}{3}30$ & $\frac{2}{3}24$ & $0$ & $30$%
\end{tabular}

Note that the first airline receives nothing. This makes sense, as no
passenger has taken any of the flights it offered. In the case of the second
airline, rule E is more favorable than rule W, since the weights of its
flights are small.
\end{example}

\bigskip

Finally, we consider the family of all weighted flights rules: 
\begin{equation*}
\mathcal{F=}\left\{ W^{w}:w\text{ is a weight system}\right\} .
\end{equation*}

Note that each weighted flights rule $W^{w}$ and the equal flights rule
belong to this family.

\bigskip

\section{Axioms}

In this section, we introduce several axioms. Some of them are inspired by
standard axioms from the literature on cost/resource allocation. Some axioms
are specific to the airlines problem.

\bigskip

Given a subset of passengers $T\subseteq M$, we can define the airlines
problem restricted to that set of passengers as $A_{|T}=\left(
N,F,T,\left(f^j,p^j\right) _{j\in T}\right)$. Note that this problem is well
defined.

\bigskip

\textbf{Additivity} is a standard axiom that says that the rule is additive
on the set of passengers; that is, distributing the benefits generated by a
group of passengers and then distributing the benefits of the remaining
group, or distributing the total benefit all at once, should yield the same
result. Formally, given a problem $A$ and a subset of passengers $T\subseteq
M$, for each $i\in N$ 
\begin{equation*}
R_i(A)=R_i(A_{|T})+R_i(A_{|M\setminus T}).
\end{equation*}

This axiom is also aligned with the procedure followed by IATA, since the
settlement of compensations among airlines is carried out on a weekly basis.
Airlines send their billing information every week, and the system
calculates how much each airline receives that week. Then, $T$ could be
considered as the set of tickets of a given week.

\bigskip

\textbf{Null airline} says that if an airline has no passengers, then it
receives nothing. This axiom is quite standard and has been used in many
problems and axiomatic characterizations. We now introduce it formally. We
say that $i$ is a null airline when $i$ has no passengers, namely, $%
M_{i}=\varnothing .$ Let $A$ be a problem and $i\in N$ a null airline. Then, 
$R_{i}\left( A\right) =0.$

\bigskip

The axiom of independence of irrelevant alternatives, introduced by Nash in
bargaining problems, says that the allocation should not depend on certain
aspects of the problem. Subsequently, many papers have applied this idea in
several different contexts. For instance, Derks and Haller (1999) introduced
the axiom of null players out, which says that removing a null player in a
cooperative game does not change the allocation to the remaining players. We
apply this same idea to airlines problems.

\textbf{Independence of empty flights} says that if a certain flight is
empty, no passenger uses it, canceling such a flight will not affect the
final allocation. Formally, consider a flight $(e,i)\in F$ such that $%
(e,i)\notin\cup _{j\in M}f^{j}$. We define the problem $\widetilde{A}=(%
\widetilde{N}, \widetilde{F},\widetilde{M},(\widetilde{f}^{j},\widetilde{p}%
^{j})_{j\in M})$, where such flight is canceled, as:

\begin{itemize}
\item $\widetilde{F}=F\setminus\{(e,i)\}$

\item $\widetilde{N}=N\setminus\{i\}$ if airline $i$ only offers the flight $%
\{(e,i)\}$ (namely $E_i =\{e \}$) and $\widetilde{N}=N$ otherwise.

\item $\widetilde{M}=M$.

\item $(\widetilde{f}^j,\widetilde{p}^j)=(f^j,p^j)$ for all $j\in \widetilde{%
M}$.
\end{itemize}

Then, $R_{i^{\prime }}(A)=R_{i^{\prime }}(\widetilde{A})$ for each $%
i^{\prime }\in \widetilde{N}$.

\bigskip

Even though null airline and independence of empty flights are
philosophically different, they are mathematically related. In fact, if a
rule $R$ satisfies independence of empty airlines, then $R$ also satisfies
null airline. We now prove this result.

Let $A$ be a problem and let $i\in N$ be a null airline in $A$. Let $\tilde{A%
}$ be the problem obtained from $A$ by canceling all flights of airline $i$.
For each $i^{\prime }\in N\backslash \left\{ i\right\} $, by independence of
empty flights, $R_{i^{\prime }}\left( \tilde{A}\right) =R_{i^{\prime}}\left(
A\right) $. Since 
\begin{equation*}
\sum_{i^{\prime }\in N\backslash \left\{ i\right\} }R_{i^{\prime }}\left( 
\tilde{A}\right) =\sum\limits_{j\in M}p^{j}=\sum_{i^{\prime }\in
N}R_{i^{\prime }}\left( A\right) ,
\end{equation*}%
we deduce that $R_{i}\left( A\right) =0.$

Nevertheless, rules exist that satisfy null airline but not independence of
empty flights. For instance, consider the rule that divides $%
\sum\limits_{j\in M}p^{j}$ among the non-null airlines proportionally to the
number of flights operated by each airline.

\bigskip

The following axioms are specific to the airlines problems considered in
this paper.

\bigskip

\textbf{Flights equivalence} says that if all passengers have booked the
same number of flights from two airlines, then the two airlines must receive
the same.

Formally, given a problem $A$, if $i,i^{\prime }\in N$ are such that for
each $j\in M$, $|f_{i}^{j}|=|f_{i^{\prime }}^{j}|$, then $%
R_{i}(A)=R_{i^{\prime }}(A)$.

\bigskip

\textbf{Independence of other airlines} says that the amount received by an
airline depends on its own flights, but not on which airlines operate the
other flights. For instance, if a flight operated by airline $i$ is
reassigned to airline $i^{\prime }$, the rest of the airlines are not
affected.

Formally, consider a function $\sigma:F \rightarrow \mathcal{N}$ that says
for each flight $(e,i)\in F$ the new airline $\sigma(e,i)$ operating it. Let
us define the problem $A^\sigma =
(N^\sigma,F^\sigma,M^\sigma,(f^{\sigma,j},p^{\sigma,j})_{j\in M})$ as
follows:

\begin{itemize}
\item $F^\sigma=\{(e,\sigma(e,i)):(e,i)\in F\}$.

\item $N^\sigma= \{ i \in \mathcal{N}: \exists (e,k)\in F \text{ with }
\sigma(e,k)=i \}$.

\item $M^\sigma=M$.

\item For each $j\in M$, $f^{\sigma,j}=\{(e,\sigma(e,k)):(e,k)\in f^j\}$ and 
$p^{\sigma,j}=p^j$.
\end{itemize}

Assume that airline $i$ offers the same flights in both problems. That is, $%
i\in N\cap N^{\sigma }$, $\sigma (e,i)=i$ for each $(e,i)\in f_{i}$, and $%
\sigma (e,k)\neq i$ for each $k\in N\setminus \{i\}$ and $(e,k)\in f_{k}$.
Then, $R_{i}(A)=R_{i}(A^{\sigma })$.

\bigskip

Suppose that passenger $j$ has paid \euro600 for the itinerary $%
\left\{\left( \left( a,b\right) ,1\right) ,\left( \left( b,c\right)
,2\right),\left( \left( c,d\right) ,3\right) \right\} $ and passenger $%
j^{\prime }$ has paid \euro800 for the itinerary $\left\{ \left( \left(
a,b\right),1\right) ,\left( \left( b,c\right) ,2\right) ,\left( \left(
c,e\right),4\right) \right\} $. Both passengers have coincided in the first
two flights, but the third one is different. Assume that the \euro600 paid
by $j$ are allocated as follows: \euro100 to airline 1, \euro200 to airline
2 and \euro300 to airline 3. How should we allocate the \euro800 paid by $%
j^{\prime }?$ \textbf{Ratio preservation} says that airline 2 must receive
twice as much as airline 1, maintaining the same ratio as for passenger $j$.
In general, whenever two airlines share flights in different itineraries,
their relative shares remain stable, regardless of the total amount paid by
the passengers or the presence of additional flights. Formally, given a
problem $A$, $j,j^{\prime }\in M$ and $i,i^{\prime }\in N^{j}\cap
N^{j^{\prime }}$ such that $f_{i}^{j}=f_{i}^{j^{\prime }}$ and $%
f_{i^{\prime}}^{j}=f_{i^{\prime }}^{j^{\prime }}$ we have that 
\begin{equation*}
\frac{R_{i}\left( A_{|\{j\}}\right) }{R_{i^{\prime }}\left(A_{|\{j\}}\right) 
}=\frac{R_{i}\left( A_{|\{j^{\prime }\}}\right) }{R_{i^{\prime }}\left(
A_{|\{j^{\prime }\}}\right) }.
\end{equation*}

\bigskip

Suppose that in the definition of a problem $A$ we also include a weight
system $w$. We now introduce an axiom related to this weight system. Berganti%
\~{n}os and Moreno-Ternero (2025a) introduced the axiom of pairwise
homogeneity for music-streaming problems. It says that if each user streams
an artist a fixed multiple of times more than another artist, the index
should preserve that proportion. We now apply the same idea to airlines
problems, using weights instead of streams.

\textbf{Pairwise homogeneity} says that if, for each passenger, the weights
of the flights operated by two airlines are in a given proportion, then the
total benefit received by these two airlines should respect the same
proportion.

Formally, given a problem $A$, $i,i^{\prime }\in N$ and $\lambda >0$ such
that for each $j\in M,$ $\sum\limits_{(e,i)\in
f_{i}^{j}}w_{e}=\lambda\sum\limits_{(e,i^{\prime })\in f_{i^{\prime
}}^{j}}w_{e}$ then, $R_{i}(A)=\lambda R_{i^{\prime }}(A).$

\bigskip

\section{The family of weighted flights rules}

We provide two characterizations of the family of weighted flights rules $%
\mathcal{F}$ by using four axioms. Three of them - additivity, independence
of other airlines and ratio preservation - are common to both
characterizations. The fourth one is either null airline or independence of
empty flights. Note that none of the axioms make any reference to the
existence of weights.

\bigskip

We now present a characterization of the family of weighted flights rule $%
\mathcal{F}$.

\bigskip

\begin{theorem}
\label{th char family} A rule satisfies additivity, null airline,
independence of other airlines, and ratio preservation if and only if $R\in $
$\mathcal{F}$.
\end{theorem}

\bigskip

\begin{proof}
It is straightforward to prove that, for any weight system $%
w=\left(w_{e}\right) _{e\in E}$, the associated weighted flights rule $W$
satisfies additivity, null airline, independence of other airlines, and
ratio preservation.

Let us consider a rule $R$ that satisfies the four axioms. By additivity, we
know that $R(A)=\sum\limits_{j\in M}R(A_{|\{j\}})$. Then, it is enough to
prove that for each $j\in M,$ $R(A_{|\{j\}})=W(A_{|\{j\}})$.

Let $j\in M.$ We first assume that each flight is operated by a different
airline. Namely, $\left\vert F\right\vert =\left\vert N\right\vert $. We
take $\left( e,i^{e}\right) \in f^{j}.$ We define $w_{e}=1.$ Let $%
\left(e^{\prime },i^{e^{\prime }}\right) \in F$ with $e^{\prime }\neq e.$
Since each pair of airports are connected in $F$ we consider the problem $%
A_{|\{j^1\}}$ where $f^{j^{1}}=$ $\left\{
\left(e_{k}^{j^{1}},i_{k}^{j^{1}}\right) \right\} _{k=1}^{n^{j^{1}}}$, $%
\left( e_{1}^{j^{1}},i_{1}^{j^{1}}\right) =\left( e,i^{e}\right) $ and $%
\left(e_{n^{j^{1}}}^{j^{1}},i_{n^{j^{1}}}^{j^{1}}\right) =\left(
e^{\prime},i^{e^{\prime }}\right) .$ We define 
\begin{equation*}
w_{e^{\prime }}=\frac{R_{i^{e^{\prime }}}\left( A_{|\{j^{1}\}}\right) }{%
R_{i^{e}}\left( A_{|\{j^{1}\}}\right) }.
\end{equation*}

We now prove that $w_{e^{\prime }}$ is well defined, namely, it does not
depend on $i^{e},$ $i^{e^{\prime }}$, and $j^{1}.$ Let $i\in \mathcal{N}$
with $i\notin N\backslash \left\{ i^{e}\right\} .$ Let $\sigma :F\rightarrow
\left( N\backslash \left\{ i^{e}\right\} \right) \cup \left\{ i\right\} $
such that $\sigma \left(e,i^{\prime }\right) =i$ when $i^{\prime e}$ and $%
\sigma \left(e, i^{\prime}\right) =i^{\prime }$ otherwise. We consider the
problem $A_{|\{j^{1}\}}^{\sigma }$ as in the definition of independence of
other airlines. By independence of other airlines, for all $i^{\prime }\in
N\backslash \left\{ i^{e}\right\} ,$ $R_{i^{\prime
}}\left(A_{|\{j^{1}\}}^{\sigma }\right) =R_{i^{\prime }}\left(
A_{|\{j^{1}\}}\right)$. Then, $R_{i}\left( A_{|\{j^{1}\}}^{\sigma }\right)
=R_{i^{e}}\left(A_{|\{j^{1}\}}\right) $ and hence $w_{e^{\prime }}$ does not
depend on $i^{e}.$ Similarly, we can argue that $w_{e^{\prime }}$ does not
depend on $i^{e^{\prime }}$.

We now prove that $w_{e^{\prime }}$ does not depend on the passenger. Let $%
f^{j^{2}}=\left\{ \left( e_{k}^{j^{2}},i_{k}^{j^{2}}\right)
\right\}_{k=1}^{n^{j^{2}}}$, $\left(e_{1}^{j^{2}},i_{1}^{j^{2}}\right)
=\left(e,i^{e}\right) $ and $%
\left(e_{n^{j^{2}}}^{j^{2}},i_{n^{j^{2}}}^{j^{2}}\right) =\left(
e^{\prime},i^{e^{\prime }}\right) .$ By ratio preservation, 
\begin{equation*}
\frac{R_{i^{e^{\prime }}}\left( A_{|\{j^{1}\}}\right) }{R_{i^{e}}\left(A_{|%
\{j^{1}\}}\right) }=\frac{R_{i^{e^{\prime }}}\left( A_{|\{j^{2}\}}\right) }{%
R_{i^{e}}\left( A_{|\{j^{2}\}}\right) }.
\end{equation*}

Thus, $w_{e^{\prime }}$ is well defined.

We now prove that $R\left( A_{|\{j\}}\right) =W\left( A_{|\{j\}}\right)$.
Let $i\in N\backslash N^{j}.$ By null airline, $R_{i}\left(A_{|\{j\}}\right)
=0=W_{i}\left( A_{|\{j\}}\right) .$

For each $i\in N^{j},$ we denote by $e^{i}$ the unique flight operated by
airline $i.$ Given $i\in N^{j}$, 
\begin{eqnarray*}
W_{i}\left( A_{|\{j\}}\right) &=&\frac{w_{e^{i}}}{\sum\limits_{i^{\prime}\in
N^{j}}w_{e^{i^{\prime }}}}p^{j}=\frac{\frac{R_{i}\left(A_{|\{j\}}\right) }{%
R_{i^{e}}\left( A_{|\{j\}}\right) }}{\sum\limits_{i^{\prime }\in N^{j}}\frac{%
R_{i^{\prime }}\left( A_{|\{j\}}\right) }{R_{i^{e}}\left( A_{|\{j\}}\right) }%
}p^{j} \\
&=&\frac{R_{i}\left( A_{|\{j\}}\right) }{\sum\limits_{i^{\prime }\in
N^{j}}R_{i^{\prime }}\left( A_{|\{j\}}\right) }p^{j}=R_{i}\left(A_{|\{j\}}%
\right) .
\end{eqnarray*}

We now consider the general case in which airlines may operate several
flights in $A_{|\{j\}}.$ Let $N^{\prime }\subset \mathcal{N}$ be such that $%
N\cap N^{\prime }=\varnothing $ and $\left\vert N^{\prime }\right\vert
=\left\vert F\right\vert .$ Let $\sigma:F\rightarrow N^{\prime }$ be a
one-to-one function. We consider the problem $A_{|\{j\}}^{\sigma }$ as in
the definition of independence of other airlines. In $A_{|\{j\}}^{\sigma }$,
each flight is operated by a different airline. Then, $R\left(
A_{|\{j\}}^{\sigma}\right) =W\left( A_{|\{j\}}^{\sigma }\right) .$

Let $i\in N.$ We consider $\sigma ^{i}$ obtained from $\sigma $ by assigning
all flights of airline $i$ in $A_{|\{j\}}$ to airline $i$. Formally, we
define $\sigma ^{i}:F\rightarrow N^{\prime }$ such that $\sigma
^{i}\left(e,i^{\prime }\right) =\sigma \left( e,i^{\prime }\right) $ when $%
i^{\prime}\neq i$ and $\sigma ^{i}\left( e,i^{\prime }\right) =i$ when $%
i^{\prime }=i$. By independence of other airlines, $R_{i}\left(A_{|\{j\}}%
\right)=R_{i}\left( A_{|\{j\}}^{\sigma ^{i}}\right) $ and $%
R_{i^{\prime}}\left(A_{|\{j\}}^{\sigma ^{i}}\right) =R_{i^{\prime
}}\left(A_{|\{j\}}^{\sigma}\right) $ when $i^{\prime }\in N^{\prime
}\backslash \left\{ i^{\prime }\in N^{\prime }:\sigma ^{-1}\left( i^{\prime
}\right) \notin f_{i}\right\} .$ Now, 
\begin{eqnarray*}
R_{i}\left( A_{|\{j\}}^{\sigma ^{i}}\right)
&=&p^{j}-\sum\limits_{i^{\prime}\in N^{\prime }:\sigma ^{-1}\left( i^{\prime
}\right) \notin f_{i}}R_{i^{\prime }}\left( A_{|\{j\}}^{\sigma ^{i}}\right)
\\
&=&p^{j}-\sum\limits_{i^{\prime }\in N^{\prime }:\sigma ^{-1}\left(i^{\prime
}\right) \notin f_{i}}R_{i^{\prime }}\left( A_{|\{j\}}^{\sigma}\right) \\
&=&p^{j}-\sum\limits_{i^{\prime }\in N^{\prime }:\sigma ^{-1}\left(i^{\prime
}\right) \notin f_{i}}W_{i^{\prime }}\left( A_{|\{j\}}^{\sigma}\right) \\
&=&p^{j}-\sum\limits_{i^{\prime }\in N\backslash \left\{
i\right\}}W_{i^{\prime }}\left( A_{|\{j\}}\right) \\
&=&W_{i}\left( A_{|\{j\}}\right) .
\end{eqnarray*}
\end{proof}

\bigskip

Since independence of empty flights implies null airline, and by Theorem \ref%
{th char family}, we obtain the following characterization of the family of
weighted flights rules.

\bigskip

\begin{corollary}
\label{cor char fam}A rule satisfies additivity, independence of empty
flights, independence of other airlines, and ratio preservation if and only
if $R$ is a weighted flights rule.
\end{corollary}

\bigskip

\bigskip

We now show that the axioms used in Theorem \ref{th char family} and
Corollary \ref{cor char fam} are independent.

\begin{itemize}
\item The rule $R^{1}$ allocates the whole benefit proportionally to the
weights of the non-empty flights of all passengers of the airline. That is,
for each $i\in N$, 
\begin{equation*}
R_{i}^{1}(A)=\dfrac{\sum\limits_{j\in M}\sum\limits_{(e,i)\in f_{i}^{j}}w_{e}%
}{\sum\limits_{i\in N}\sum\limits_{j\in M}\sum\limits_{(e,i)\in
f_{i}^{j}}w_{e}}\sum_{j\in M}p^{j}.
\end{equation*}%
$R^{1}$ satisfies null airline, independence of empty flights, independence
of other airlines, and ratio preservation, but fails additivity.

\item The rule $R^{2}$ divides the price paid by each passenger equally
among the airlines not used by that passenger. Formally, given $j\in M$ we
define $R^{2}(A_{|\{j\}})$ as follows. If $N^{j}\neq N$ we define $%
R_{i}^{2}\left( A_{|\{j\}}\right) =0$ if $i\in N^{j}$ and $R_{i}^{2}\left(
A_{|\{j\}}\right) =\frac{p^j}{\left\vert N\backslash Nj\right\vert }$ if $%
i\notin N^{j}.$ If $N^{j}=N$ we define $R^{2}\left(A_{|\{j\}}\right)
=W\left( A_{|\{j\}}\right) .$ Now, $R^{2}\left( A\right) =\sum\limits_{j\in
M}R^{2}\left( A_{|\{j\}}\right) .$

$R^{2}$ satisfies additivity, independence of other airlines, and ratio
preservation, but fails null airline and independence of empty flights.

\item The rule $R^{3}$ divides the price paid by each passenger equally
among the airlines that operate at least one of the passenger's flights. That is, 
\begin{equation*}
R_{i}^{3}(A)=\left\{ 
\begin{array}{cl}
0 & \text{if }i\notin \cup _{j\in M}N^{j} \\ 
\sum\limits_{j\in M:i\in N^{j}}\frac{p^{j}}{|N^{j}|} & \text{otherwise}%
\end{array}
\right.
\end{equation*}
$R^{3}$ satisfies additivity, null airline, independence of empty flights,
and ratio preservation, but fails independence of other airlines.

\item The rule $R^{4}$ divides the price paid by each passenger $j$
following a weighted flights rule whose weights depend on passenger $j$.
Formally, for each problem $A$ and each $i\in N,$ 
\begin{equation*}
R_{i}^{3}\left( A\right) =\sum_{j\in M}\frac{\sum\limits_{(e,i)\in
f_{i}^{j}}w_{e}^{j}}{\sum\limits_{(e,k)\in f^{j}}w_{e}^{j}}p^{j}.
\end{equation*}

$R^{4}$ satisfies additivity, null airline, independence of empty flights,
and independence of other airlines, but fails ratio preservation.
\end{itemize}

\bigskip

In this characterization, there is no axiom related to weights. Theorem \ref%
{th char family} says that if we want a rule satisfying the four axioms,
then we must use a weighted flights rule. Thus, we should compute the
\textquotedblleft appropriate weights\textquotedblright\ and then apply the
associated weighted rule. Hence, this theorem provides theoretical support
for the approach used by IATA.

\bigskip

\section{The weighted flights rule}

In this section, we assume that the definition of a problem $A$ also
includes a weight system $w$. Thus, it makes sense to consider the axiom of
pairwise homogeneity. We present three characterizations of the weighted
flights rule. Two axioms - additivity and pairwise homogeneity - appear in
the three characterizations, while the other three - null airline,
independence of empty flights, and independence of other airlines - each
appear in only one.

\bigskip

In the next proposition, whose straightforward proof is omitted, we discuss
the axioms satisfied by the weighted flights rule.

\bigskip

\begin{proposition}
\label{axioms weighted rule}The weighted flights rule satisfies additivity,
null airline, independence of empty flights, independence of other airlines,
ratio preservation, and pairwise homogeneity.

The weighted flights rule fails flights equivalence.
\end{proposition}

\bigskip

In our first result, we give a characterization of the weighted flights rule
on the whole domain.

\bigskip

\begin{theorem}
\label{th char weighted} The weighted flights rule is the only rule
satisfying additivity, null airline, and pairwise homogeneity.
\end{theorem}

\bigskip

\begin{proof}
$\left( a\right) $ By Proposition \ref{axioms weighted rule}, $W$ satisfies
the three axioms.

Let us consider a rule $R$ that satisfies the three axioms. By additivity,
it is enough to prove that $R(A_{|\{j\}})=W(A_{|\{j\}})$ for all $j\in M$.

Let $j\in M.$ Let $i\in N$ be such that $i\notin N^{j}.$ By null airlines, $%
R_{i}(A_{|\{j\}})=0=W_{i}(A_{|\{j\}}).$

Let $i\in N$ be such that $i\in N^{j}.$ If $N^{j}=\left\{ i\right\} ,$ then 
\begin{equation*}
R_{i}(A_{|\{j\}})=p^{j}-\sum\limits_{i^{\prime }\in N\backslash
\left\{i\right\} }R_{i}\left( A_{|\{j\}}\right)
=p^{j}=W_{i}\left(A_{|\{j\}}\right) .
\end{equation*}

Assume that $N^{j}\neq \left\{ i\right\} .$ Let $i^{\prime }\in
N^{j}\backslash \left\{ i\right\} .$ Since $\sum\limits_{(e,i^{\prime })\in
f_{i^{\prime }}^{j}}w_{e}=\lambda \sum\limits_{(e,i)\in f_{i}^{j}}w_{e}$
where $\lambda =\frac{\sum\limits_{(e,i^{\prime })\in
f_{i^{\prime}}^{j}}w_{e}}{\sum\limits_{(e,i)\in f_{i}^{j}}w_{e}},$ by
pairwise homogeneity, $R_{i^{\prime }}(A_{|\{j\}})=\lambda R_{i}(A_{|\{j\}}).
$

Then, 
\begin{eqnarray*}
p^{j} &=&\sum\limits_{i^{\prime }\in
N^{j}}R_{i^{\prime}}(A_{|\{j\}})=\sum\limits_{i^{\prime }\in N^{j}\backslash
\left\{ i\right\} }\frac{\sum\limits_{(e,i^{\prime })\in f_{i^{\prime
}}^{j}}w_{e}}{\sum\limits_{(e,i)\in f_{i}^{j}}w_{e}}R_{i}(A_{|\{j%
\}})+R_{i}(A_{|\{j\}})\Rightarrow \\
R_{i}(A_{|\{j\}}) &=&\frac{\sum\limits_{(e,i)\in f_{i}^{j}}w_{e}}{%
\sum\limits_{i^{\prime }\in N^{j}}\sum\limits_{(e,i^{\prime })\in
f_{i^{\prime }}^{j}}w_{e}}p^{j}=W_{i}\left( A_{|\{j\}}\right) .
\end{eqnarray*}
\end{proof}

\bigskip

Since independence of empty flights implies null airline, by Proposition \ref%
{axioms weighted rule} and Theorem \ref{th char weighted} we have the
following characterization of the weighted flights rule.

\bigskip

\begin{corollary}
\label{cor char weihted} The weighted flights rule is the only rule
satisfying additivity, independence of empty flights, and pairwise
homogeneity.
\end{corollary}

\bigskip

We now argue that the axioms used in Theorem \ref{th char weighted} and
Corollary \ref{cor char weihted} are independent.

\begin{itemize}
\item The rule $R^{1}$ defined above satisfies null airline, independence of
empty flights, and pairwise homogeneity, but fails additivity.

\item The equal flights rule satisfies additivity, null airline, and
independence of empty flights, but fails pairwise homogeneity.

\item The rule $R^{2}$ defined above satisfies additivity and pairwise
homogeneity, but fails null airline and independence of empty flights.
\end{itemize}

\medskip

\bigskip

We now consider the subdomain of airlines problems where each passenger has
taken at least two flights. This is the relevant domain in reality because
if a passenger only takes one flight, then the price paid by the passenger
goes to the airline operating the flight.

\bigskip

\begin{theorem}
\label{th char weighted subdomain} When $\left\vert f^{j}\right\vert \geq 2$
for all $j\in M,$ the weighted flights rule is the only rule satisfying
additivity, independence of other airlines, and pairwise homogeneity.
\end{theorem}

\bigskip

\begin{proof}
By Proposition \ref{axioms weighted rule}, $W$ satisfies the three axioms.

Let us consider a rule $R$ that satisfies the three axioms. By additivity,
as in the previous theorem, it is enough to prove that $R(A_{|\{j\}})=W(A_{|%
\{j\}})$ for all $j\in M$.

Let $j\in M.$ We consider $N^{\prime }\subset \mathcal{N}$ such that $N\cap
N^{\prime }=\varnothing $ and $\left\vert N^{\prime }\right\vert =\left\vert
F\right\vert .$ Let $\sigma :F\rightarrow N^{\prime }$ be a one-to-one
function. We consider the problem $A_{|\{j\}}^{\sigma }$ as in the
definition of independence of other airlines.

Let $i,i^{\prime }\in N^{\prime }$ be such that $\left\{ \sigma
^{-1}\left(i\right) ,\sigma ^{-1}\left( i^{\prime }\right) \right\} \subset
f^{j}$. By pairwise homogeneity, $R_{i}\left( A_{|\{j\}}^{\sigma }\right) =%
\frac{w_{\sigma ^{-1}\left( i\right) }}{w_{\sigma ^{-1}\left( i^{\prime
}\right) }}R_{i^{\prime }}\left( A_{|\{j\}}^{\sigma }\right) \footnote{%
We make an abuse of notation. $\sigma ^{-1}\left( i\right) =\left(
e^{\ast},i^{\ast }\right) \in f^{j}$. We write $w_{\sigma ^{-1}\left(
i\right) }$ instead of $w_{e^{\ast }}.$}.$ Let $x=\sum\limits_{\sigma
^{-1}\left(i\right) \in f^{j}}R_{i}\left( A_{|\{j\}}^{\sigma }\right) .$
Then, for all $i\in N^{\prime }$ such that $\sigma ^{-1}\left( i\right) \in
f^{j},$ 
\begin{equation*}
R_{i}\left( A_{|\{j\}}^{\sigma }\right) =\frac{w_{\sigma ^{-1}\left(i\right)
}}{\sum\limits_{\sigma ^{-1}\left( i^{\prime }\right) \in f^{j}}w_{\sigma
^{-1}\left( i^{\prime }\right) }}x.
\end{equation*}

We now prove that for each $i^{\prime }\in N^{\prime }$ such that $%
\sigma^{-1}\left( i^{\prime }\right) \notin f^{j},$ $R_{i^{\prime
}}\left(A_{|\{j\}}^{\sigma }\right) =0.$ Let $i^{\ast },i^{\prime }\in
N^{\prime }$ be such that $\sigma ^{-1}\left( i^{\ast }\right) \in f^{j}$
and $\sigma^{-1}\left( i^{\prime }\right) \notin f^{j}.$ We now consider $%
\sigma^{\prime }$ obtained from $\sigma $ by reassigning the flight of $%
i^{\prime} $ to $i^{\ast }.$ Formally, we define $\sigma ^{\prime
}:F\rightarrow N^{\prime }$ such that $\sigma ^{\prime }\left( e,i\right)
=i^{\ast }$ when $\left( e,i\right) =\sigma ^{-1}\left( i^{\prime }\right) $
and $\sigma^{\prime }\left( e,i\right) =\sigma \left( e,i\right) $ otherwise.

By independence of other airlines, for all $i\in N^{\prime }\backslash
\left\{ i^{\ast },i^{\prime }\right\} $ $R_{i}\left(
A_{|\{j\}}^{\sigma}\right) =R_{i}\left( A_{|\{j\}}^{\sigma ^{\prime
}}\right) .$ Then, 
\begin{eqnarray*}
R_{i^{\ast }}\left( A_{|\{j\}}^{\sigma ^{\prime
}}\right)&=&p^{j}-\sum\limits_{i\in i\in N^{\prime }\backslash \left\{
i^{\ast},i^{\prime }\right\} }R_{i}\left( A_{|\{j\}}^{\sigma ^{\prime
}}\right) \\
&=&p^{j}-\sum\limits_{i\in i\in N^{\prime }\backslash \left\{
i^{\ast},i^{\prime }\right\} }R_{i}\left( A_{|\{j\}}^{\sigma }\right) \\
&=&R_{i^{\ast }}\left( A_{|\{j\}}^{\sigma }\right) +R_{i^{\prime
}}\left(A_{|\{j\}}^{\sigma }\right) .
\end{eqnarray*}

Since $\left\vert f^{j}\right\vert \geq 2,$ $\left\vert
N^{\prime}\right\vert \geq 2.$ Then, we can find $i^{\ast \ast }\in
N^{\prime}\backslash \left\{ i^{\ast }\right\} $ such that $\sigma
^{-1}\left(i^{\ast \ast }\right) \in f^{j}.$ By pairwise homogeneity applied
to airlines $i^{\ast }$ and $i^{\ast \ast }$ in $A_{|\{j\}}^{\sigma ^{\prime
}}$ we have that 
\begin{eqnarray*}
R_{i^{\ast }}\left( A_{|\{j\}}^{\sigma ^{\prime }}\right) &=&\frac{%
w_{\sigma^{-1}\left( i^{\ast }\right) }}{w_{\sigma ^{-1}\left( i^{\ast \ast
}\right) }}R_{i^{\ast \ast }}\left( A_{|\{j\}}^{\sigma ^{\prime }}\right) =%
\frac{w_{\sigma ^{-1}\left( i^{\ast }\right) }}{w_{\sigma ^{-1}\left(
i^{\ast \ast}\right) }}R_{i^{\ast \ast }}\left( A_{|\{j\}}^{\sigma }\right)
\\
&=&\frac{w_{\sigma ^{-1}\left( i^{\ast }\right) }}{w_{\sigma
^{-1}\left(i^{\ast \ast }\right) }}\frac{w_{\sigma ^{-1}\left( i^{\ast \ast
}\right) }}{\sum\limits_{\sigma ^{-1}\left( i^{\prime }\right) \in
f^{j}}w_{\sigma^{-1}\left( i^{\prime }\right) }}x=\frac{w_{\sigma
^{-1}\left( i^{\ast}\right) }}{\sum\limits_{\sigma ^{-1}\left( i^{\prime
}\right) \in f^{j}}w_{\sigma ^{-1}\left( i^{\prime }\right) }}x \\
&=&R_{i^{\ast }}\left( A_{|\{j\}}^{\sigma }\right) .
\end{eqnarray*}

Then, $R_{i^{\prime }}\left( A_{|\{j\}}^{\sigma }\right) =0$.

Hence, $x=p^{j},$ and we conclude that $R\left( A_{|\{j\}}^{\sigma
}\right)=W\left( A_{|\{j\}}^{\sigma }\right) .$

We now prove that for each $i^{\prime }\in N,$ $R_{i}(A_{|\{j\}})=W_{i}(A_{|%
\{j\}})$. We define $\sigma ^{i}$ from $\sigma $ by reassigning to airline $i
$ its flights in $A_{|\{j\}}.$ Formally, we define $\sigma^{i}:F\rightarrow
N^{\prime }$ such that $\sigma ^{i}\left( e,i^{\prime}\right) =\sigma \left(
e,i^{\prime }\right) $ when $i^{\prime}\neq i$ and $\sigma ^{i}\left(
e,i^{\prime }\right) =i$ when $i^{\prime }=i$. By independence of other
airlines, $R_{i^{\prime }}\left( A_{|\{j\}}^{\sigma }\right) =R_{i^{\prime
}}\left( A_{|\{j\}}^{\sigma ^{i}}\right) $ when $i^{\prime }\in N^{\prime
}\backslash \left\{ i^{\prime }\in N^{\prime}:\sigma ^{-1}\left( i^{\prime
}\right) \notin f_{i}\right\}$. Then, 
\begin{eqnarray*}
R_{i}\left( A_{|\{j\}}^{\sigma ^{i}}\right) &=&p^{j}-\sum\limits_{i^{\prime
}\in N^{\prime }:\sigma ^{-1}\left( i^{\prime }\right) \notin f_{i}}
R_{i^{\prime }}\left( A_{|\{j\}}^{\sigma ^{i}}\right)
=p^{j}-\sum\limits_{i^{\prime }\in N^{\prime }:\sigma ^{-1}\left(i^{\prime
}\right) \notin f_{i}}R_{i^{\prime }}\left(A_{|\{j\}}^{\sigma }\right) \\
&=&p^{j}-\sum\limits_{i^{\prime }\in N^{\prime }:\sigma ^{-1}\left(
i^{\prime }\right) \notin f_{i}}W_{i^{\prime }}\left( A_{|\{j\}}^{\sigma
}\right)=p^{j}-\sum\limits_{i^{\prime }\in N\backslash \left\{
i\right\}}W_{i^{\prime }}\left( A_{|\{j\}}\right) \\
&=&W_{i}\left( A_{|\{j\}}\right) .
\end{eqnarray*}%
By independence of other airlines, we have that $R_{i}\left(A_{|\{j\}}^{%
\sigma ^{i}}\right) =R_{i}\left( A_{|\{j\}}\right) .$
\end{proof}

\bigskip

We now argue that the axioms used in Theorem \ref{th char weighted subdomain}
are independent.

\begin{itemize}
\item $R^{1}$ defined above satisfies all axioms but additivity.

\item The equal flights rule satisfies all axioms but pairwise homogeneity.

\item $R^{2}$ defined above satisfies all axioms but null airline.
\end{itemize}

\bigskip

\section{The equal flights rule}

In this section, we consider that there is no weight system associated with
the problem. We study the equal flights rule. The main result of the section
is a characterization based on three axioms: additivity, flights
equivalence, and independence of other airlines.

\bigskip

In the next proposition, whose straightforward proof is omitted, we discuss
the axioms satisfied by the equal flights rule.

\bigskip

\begin{proposition}
\label{axioms equal rule}The equal flights rule satisfies additivity, null
airline, independence of empty flights, flights equivalence, independence of
other airlines, and ratio preservation.
\end{proposition}

\bigskip

Since the definition of the problem does not include a weight system, the
axiom of pairwise homogeneity cannot be applied.

\bigskip

We now provide a characterization of the equal flights rule under the
assumption that $\left\vert F\backslash f^{j}\right\vert \geq 3$ for each $%
j\in M$; that is, for each passenger, there are at least three flights not
taken by that passenger. This assumption is clearly realistic.

\begin{theorem}
\label{th char equal subdomain} When $\left\vert F\backslash
f^{j}\right\vert \geq 3$ for each $j\in M,$ the equal flights rule is the
only rule satisfying additivity, flights equivalence and independence of
other airlines.
\end{theorem}

\bigskip

\begin{proof}
By Proposition \ref{axioms equal rule}, the equal flights rule satisfies the
three axioms.

Let us consider a rule $R$ that satisfies the three axioms. By additivity,
we know that $R(A)=\sum\limits_{j\in M}R(A_{|\{j\}})$. Then, it is enough to
prove that for each $j\in M,$ $R(A_{|\{j\}})=E(A_{|\{j\}})$.

Let $j\in M.$ We consider $N^{\prime }\subset \mathcal{N}$ such that $N\cap
N^{\prime }=\varnothing $ and $\left\vert N^{\prime }\right\vert =\left\vert
F\right\vert .$ Let $\sigma :F\rightarrow N^{\prime }$ be a one-to-one
function. We consider the problem $A_{|\{j\}}^{\sigma }$ as in the
definition of independence of other airlines.

Let $i,i^{\prime }\in N^{\prime }$ be such that $\left\{ \sigma
^{-1}\left(i\right) ,\sigma ^{-1}\left( i^{\prime }\right) \right\} \subset
f^{j}$. By flights equivalence, $R_{i}\left( A_{|\{j\}}^{\sigma }\right)
=R_{i^{\prime }}\left( A_{|\{j\}}^{\sigma }\right) =x.$ Let $i,i^{\prime
}\in N^{\prime }$ be such that $\left\{ \sigma ^{-1}\left( i\right) ,\sigma
^{-1}\left(i^{\prime }\right) \right\} \cap f^{j}=\varnothing $. Again, by
flights equivalence, $R_{i}\left( A_{|\{j\}}^{\sigma }\right) =R_{i^{\prime
}}\left(A_{|\{j\}}^{\sigma }\right) =y.$ Note that $p^{j}=\left\vert
f^{j}\right\vert x+\left\vert F\backslash f^{j}\right\vert y.$

Let $i^{1}\in N^{\prime }$ be such that $\sigma ^{-1}\left(
i^{1}\right)\notin f^{j}.$ Since $\left\vert F\backslash f^{j}\right\vert
\geq 3,$ $\left\vert \left\{ i^{\prime }\in N^{\prime }:\sigma ^{-1}\left(
i^{\prime}\right) \notin f^{j}\right\} \right\vert \geq 3.$ Let $i^{2}\in
N^{\prime }$ be such that $\sigma ^{-1}\left( i^{2}\right) \notin f^{j}$ and 
$i^{1}\neq i^{2}.$

We now consider $\sigma ^{\prime }$ where all flights not taken by $j$ but
the one associated with $i^{1}\ $are reassigned to $i^{2}.$ Formally, we
define $\sigma ^{\prime }:F\rightarrow N^{\prime }$ such that $%
\sigma^{\prime }\left( e,i\right) =\sigma \left( e,i\right) $ when $%
\left(e,i\right) \in f^{j}$ or $\sigma \left( e,i\right) =i^{1}$ and $%
\sigma^{\prime }\left( e,i\right) =i^{2}$ otherwise.

By independence of other airlines, for all $i\in \left\{ i^{\prime }\in
N^{\prime }:\sigma ^{-1}\left( i^{\prime }\right) \in f^{j}\right\} \cup
\left\{ i^{1}\right\} ,$ $R_{i}\left( A_{|\{j\}}^{\sigma
}\right)=R_{i}\left( A_{|\{j\}}^{\sigma ^{\prime }}\right) .$ Then, $%
R_{i^{1}}\left(A_{|\{j\}}^{\sigma ^{\prime }}\right) =y$ and 
\begin{eqnarray*}
R_{i^{2}}\left( A_{|\{j\}}^{\sigma ^{\prime
}}\right)&=&p^{j}-\sum\limits_{i\in N^{\prime }:\sigma ^{-1}\left( i\right)
\in f^{j}}R_{i}\left( A_{|\{j\}}^{\sigma ^{\prime }}\right)
-R_{i^{1}}\left(A_{|\{j\}}^{\sigma ^{\prime }}\right) \\
&=&p^{j}-\sum\limits_{i\in N^{\prime }:\sigma ^{-1}\left( i\right) \in
f^{j}}R_{i}\left( A_{|\{j\}}^{\sigma }\right)
-R_{i^{1}}\left(A_{|\{j\}}^{\sigma }\right) \\
&=&\left( \left\vert F\backslash f^{j}\right\vert -1\right) y.
\end{eqnarray*}

By flights equivalence, $R_{i^{1}}\left( A_{|\{j\}}^{\sigma
^{\prime}}\right) =R_{i^{2}}\left( A_{|\{j\}}^{\sigma ^{\prime }}\right) .$
Then, $y=\left( \left\vert F\backslash f^{j}\right\vert -1\right) y$. Since $%
\left\vert F\backslash f^{j}\right\vert \geq 3,$ $y=0.$ Thus, $x=\frac{p^{j}%
}{\left\vert f^{j}\right\vert }$.

Let $i\in N.$ We consider $\sigma ^{i}$ obtained from $\sigma $ by assigning
all flights of airline $i$ in $A_{|\{j\}}$ to airline $i.$ Formally, we
define $\sigma ^{i}:F\rightarrow N^{\prime }$ such that $\sigma
^{i}\left(e,i^{\prime }\right) =\sigma \left( e,i^{\prime }\right) $ when $%
i^{\prime}\neq i$ and $\sigma ^{i}\left( e,i^{\prime }\right) =i$ when $%
i^{\prime }=i$. By independence of other airlines, $R_{i}\left(
A_{|\{j\}}\right)=R_{i}\left( A_{|\{j\}}^{\sigma ^{i}}\right) $ and $%
R_{i^{\prime }}\left(A_{|\{j\}}^{\sigma ^{i}}\right) =R_{i^{\prime }}\left(
A_{|\{j\}}^{\sigma}\right) $ when $i^{\prime }\in N^{\prime }\backslash
\left\{ i^{\prime }\in N^{\prime }:\sigma ^{-1}\left( i^{\prime }\right)
\notin f_{i}\right\} $. Now, 
\begin{eqnarray*}
R_{i}\left( A_{|\{j\}}^{\sigma ^{i}}\right)
&=&p^{j}-\sum\limits_{i^{\prime}\in N^{\prime }:\sigma ^{-1}\left( i^{\prime
}\right) \notin f_{i}}R_{i^{\prime }}\left( A_{|\{j\}}^{\sigma ^{i}}\right)
\\
&=&p^{j}-\sum\limits_{i^{\prime }\in N^{\prime }:\sigma ^{-1}\left(i^{\prime
}\right) \notin f_{i}}R_{i^{\prime }}\left( A_{|\{j\}}^{\sigma }\right) \\
&=&p^{j}-\frac{\left\vert f^{j}-f_{i}^{j}\right\vert }{\left\vert
f^{j}\right\vert }p^{j}=\frac{\left\vert f_{i}^{j}\right\vert }{\left\vert
f^{j}\right\vert }p^{j} \\
&=&E_{i}\left( A_{|\{j\}}\right) .
\end{eqnarray*}
\end{proof}

\bigskip

We now show that the axioms used in Theorem \ref{th char equal subdomain}
are independent.

\begin{itemize}
\item The rule $R^{5}$ divides the whole price proportionally to the number
of non-empty flights operated by each company. That is, 
\begin{equation*}
R_{i}^{4}(A)=\dfrac{\sum\limits_{j\in M}|f_{i}^{j}|}{\sum\limits_{i\in
N}\sum\limits_{j\in M}|f_{i}^{j}|}\sum_{j\in M}p^{j}.
\end{equation*}%
$R^{5}$ satisfies all axioms but additivity.

\item The weighted flights rule satisfies all axioms but flights equivalence.

\item $R^{3}$ defined above satisfies all axioms but independence of other
airlines.
\end{itemize}

\bigskip

In Theorem \ref{th char weighted subdomain}, we have characterized the
weighted flights rule using additivity, independence of other airlines, and
pairwise homogeneity. In Theorem \ref{th char equal subdomain}, we have
characterized the equal flights rule by replacing pairwise homogeneity in
Theorem \ref{th char weighted subdomain} with flights equivalence. A natural
question that arises is what happens if, in Theorem \ref{th char weighted},
we also replace pairwise homogeneity with flights equivalence. The answer is
that this substitution does not lead to a characterization of the equal
flights rule, since additional rules (for instance, $R^{3}$) also satisfy
the resulting set of axioms.

\bigskip

\section{Conclusions}

We have analyzed, through an axiomatic approach, the proration mechanism
used by IATA for sharing the revenue associated with a passenger's journey.
Theorem \ref{th char family} and Corollary \ref{cor char fam} lend support
to the use of a weighted rule, as IATA does, for sharing the revenue.
Nevertheless, they do not specify which weights should be used. We then
assume that the problem has an associated weight system for measuring the
relative importance of the flights. Now, Theorem \ref{th char weighted},
Corollary \ref{cor char weihted}, and Theorem \ref{th char weighted
subdomain}, provide further support for the use of IATA's mechanism. We also
propose a way to share the revenues when we consider that all flights are
rather similar.

The table below summarizes our results. The superscripts indicate the
axiomatic characterizations. Pairwise homogeneity only applies when the
model includes weights. Since the family of the weighted flights rule and
the equal flights rule are defined in a model without weights, pairwise
homogeneity cannot be checked.

\begin{equation*}
\begin{tabular}{|l|l|l|l|}
\hline
Axioms & Family W.F.R. & Weighted F.R. & Equal F.R. \\ \hline
Additivity & Yes$^{1,2}$ & Yes$^{3,4,5}$ & Yes$^{6}$ \\ \hline
Null airline & Yes$^{1}$ & Yes$^{3}$ & Yes \\ \hline
Ind. Empty flights & Yes$^{2}$ & Yes$^{4}$ & Yes \\ \hline
Flights equiv. & No & No & Yes$^{6}$ \\ \hline
Ind. other airlines & Yes$^{1,2}$ & Yes$^{5}$ & Yes$^{6}$ \\ \hline
Ratio preserv. & Yes$^{1,2}$ & Yes & Yes \\ \hline
Pairwise hom. &  & Yes$^{3,4,5}$ &  \\ \hline
\end{tabular}%
\end{equation*}

\bigskip

Kimms and Cetiner (2012) and Wang (2020) studied the problem using
cooperative game theory. Both papers associate two different cooperative
games with the problem and study two different cooperative solution concepts
(the nucleolus and the Shapley value). The game proposed by Kimms and
Cetiner (2012) is defined taking into account some specific elements of the
problem. The game proposed by Wang (2020) is defined by applying the
classical pessimistic approach (see, for instance, Ginsburgh and Zang (2003)
and Berganti\~{n}os and Moreno-Ternero (2026)). It could be interesting to
compute the core of this pessimistic game. In fact, we have obtained some
results on this topic. The Shapley value corresponds to the rule $R^{3}$
defined in this paper and the game is convex. Hence, the core is non-empty
and the Shapley value belongs to it.

Other ways exist to define cooperative games, such as the optimistic
approach (Atay and Trudeau, 2026). It would be interesting to define other
cooperative games and to study their associated solutions.

\bigskip

Calleja $et$ $al$ (2005) introduced multi-issue allocation situations.
Subsequently, several authors have applied this model to several real-life
problems. Recently, Acosta $et$ $al$ (2023) studied water policies and
Berganti\~{n}os and Moreno-Ternero (2026) the music streaming industry. It
could be interesting to apply this framework to the problem studied in this
paper.

\end{document}